\documentclass{article}
\usepackage{graphicx} % Required for inserting images
\usepackage{algorithm}
\usepackage{algpseudocode}
\usepackage{makecell}
\usepackage{amsmath}

\title{Fast Spot Order Optimization to Increase Dose Rates in Scanned Particle Therapy FLASH Treatments}
\author{Viktor Wase$^1$, Oscar Widenfalk$^1$\footnote{Mr. Widenfalk performed his work as a master thesis worker at RaySearch Laboratories, and has since started working for Elekta.}, Rasmus Nilsson$^1$, \\ Claes Fälth$^1$\footnote{Mr. Fälth contributed to the article while being employed at RaySearch Laboratories, and has since started working for Netlight.}, Albin Fredriksson$^1$.}
\date{$^1$RaySearch Laboratories\\%
September 2024}

\begin{document}

\maketitle

\begin{abstract}
The advent of ultra-high dose rate irradiation, known as FLASH radiation therapy, has shown promising potential in reducing toxicity while maintaining tumor control. However, the clinical translation of these benefits necessitates efficient treatment planning strategies. This study introduces a novel approach to optimize proton therapy for FLASH effects using traveling salesperson problem (TSP) heuristics. We applied these heuristics to optimize the arrangement of proton spots in treatment plans for 26 prostate cancer patients, comparing the performance against conventional sorting methods and global optimization techniques. Our results demonstrate that TSP-based heuristics significantly enhance FLASH coverage to the same extent as the global optimization technique, but with computation times reduced from hours to a few seconds. This approach offers a practical and scalable solution for enhancing the effectiveness of FLASH therapy, paving the way for more effective and personalized cancer treatments. Future work will focus on further optimizing run times and validating these methods in clinical settings.
\end{abstract}

\section{Introduction}
Ultra-high dose rate irradiation, known as FLASH radiation therapy, has recently attracted substantial attention due to its potential to improve the sparing of healthy tissues while maintaining tumor control. The beneficial effects, which have been observed in preclinical studies in mice~\cite{cancers13051012, MONTAYGRUEL2017365}, cats, and mini-pigs~\cite{vozenin2019advantage}, are collectively called the ``FLASH effect''. Efficient treatment planning strategies are crucial to facilitate further investigation of the effect and the potential translation of the findings into clinical practice.

Various dose rate metrics have been proposed to determine whether a given voxel receives dose at FLASH dose rate~\cite{NonpercentileDoseRateDef, doserateDef, doserateDefPoster}. Given such a metric, a critical factor in FLASH treatments using pencil beam scanning (PBS) is the spatial and temporal distribution of the spots in the treatment plan. An optimal order of spots can maximize the volume receiving FLASH dose rates. There exists a wide literature on spot pattern optimization to speed up treatment delivery~\cite{liang2022investigation, LI2015547, DIAS2015130} or improve continuous line delivery \cite{Liu_2023}, but we were only able to find a two peer reviewed articles that applied a similar concept to achieve FLASH dose rates, by Santo et al.~\cite{genetic} and Zhao et al.\cite{ZHAO2024}. They both employed genetic algorithms to optimize the order of spots to achieve FLASH dose rates in as many voxels as possible. While the resulting quality was high, this type of algorithm often suffers from long computational run times.

In this study, we present a novel spot order optimization approach to optimize FLASH treatments using proton PBS. It is a continuation of the 2023 Master thesis by coauthor Widenfalk \cite{widenfalk2023proton}. We harness the power of traveling salesperson problem (TSP) heuristics, adapted to take dose rate measures into account, to efficiently optimize the spot order in a way that achieves FLASH dose rates in as many voxels as possible. To evaluate the proposed method, we apply the TSP heuristics to 26 prostate cases treated with proton therapy. We compare the heuristics to global optimization techniques and to conventional spot sorting strategies, both in terms of sorting time and in terms of FLASH coverage. The reduced run times of the proposed method compared to global optimization, and the increased FLASH coverage compared to conventional sorting strategies, can make FLASH treatments using PBS more accessible and clinically viable.

\section{Method}

\subsection{FLASH Dose Rate}
One of the more common measures of FLASH dose rate is the Dose-Averaged Dose Rate (DADR) \cite{dadr}, which unfortunately is independent of the order of the spots, which is why the FLASH dose rate model employed in this work was the percentile dose rate~\cite{NonpercentileDoseRateDef, doserateDef, doserateDefPoster} with a 5\% buffer. It defines the dose rate $r$ of a voxel as
$$
r := \frac{0.9 d}{t_{95\%} - t_{5\%}}
$$
where $d$ is the dose to the voxel, $t_{5\%}$ and $t_{95\%}$ are the times when 5\% and 95\% of that dose has been delivered, respectively. Note that $t_{5\%}$ and $t_{95\%}$ depend on the order of the spots of the plan, while $d$ is independent of the order. We assume constant dose rate from the beam during spot irradiation and can thus use linear interpolation between the start time and end time of the spot irradiation in order to calculate the exact value of $t_{5\%}$ and $t_{95\%}$. A voxel is said to fulfill the FLASH criterion if $r \geq 40$ Gy/s and $d \geq 4$ Gy. These values are somewhat arbitrary but based on the current understanding of the FLASH effect~\cite{10.1158/1078-0432.CCR-19-1440, cancers16040798, genetic}.

The methods presented in the present paper are agnostic regarding the choice of dose rate model and the dose and dose rate limits, but we chose to focus on a single model to reduce the scope of the study. Deffet et al.~\cite{doserateDef} define five dose rate models, two of which do not depend on the spot order and are therefore not affected by spot order optimization. The remaining three are called PBS window, percentile window and max percentile window, with percentile window being the one described above. PBS window works similarly, but using absolute dose thresholds instead of dose thresholds relative to the maximum dose. The max percentile window finds the shortest time window that fulfills the constraint that it covers $p$\% of the total dose (where $p$ is a free parameter), and then it calculates the dose rate based on that.

The percentile window model was chosen due to its short computational time and its limited parameters while still being dependant on the spot ordering.

\subsection{Spot Ordering}
%Since the percentile dose rate depends on the time at which each voxel has received certain percentages of its dose, it is affected by the order in which the spots are delivered.
In conventional PBS, the spots can be ordered in a way that allows for as fast a delivery as possible. Finding the shortest path that visits each spot position once is equivalent to the Hamiltonian path problem, and good solutions can often be found by heuristics for TSP. Here, each position to visit is called a ``node''. Heuristics for solving the TSP typically maintain at least one tour (with tour defined as an order in which to visit the nodes), and tries to iteratively improve on the tour by applying small changes to the tour, and retaining modifications that lead to improvements.

We take inspiration from existing TSP heuristics that aim to minimize the tour length (see, e.g., Johnson et al.~\cite{johnson1997traveling} and Helsgaun~\cite{helsgaun2000effective}), but modify the objective to measure the volume achieving the FLASH effect instead of tour length. Here, we use the FLASH dose rate model described in the previous subsection, but the spot ordering methods described do not depend on the specific dose rate model and could be used with others.

We employ two TSP heuristics that are both based on a simple operation with two nodes, $n_1$ and $n_2$, as input: the \textit{2-opt} and the \textit{place after}. The 2-opt heuristic is a basic but fast heuristic for finding good tours for TSP. For a proposed TSP tour, it tries to exchange two edges. This is equivalent to selecting a subtour and reversing it. Its simple operation is thus to reverse the subtour between $n_1$ and $n_2$. The \textit{place after} heuristic simply places $n_1$ after $n_2$ in the tour. In initial experiments, we also employed the \textit{3-opt} heuristic which exchanges three edges, but it was found to lead to longer computational times without providing any substantial benefit, and was hence dropped.

We run both TSP heuristics for all combinations of $n_1$ and $n_2$, and we stop the process after no improvement has been found. The process is parallelized over $n_2$. The procedure for running a TSP heuristic of this type is detailed in Algorithm \ref{alg:heur}. The initial spot ordering is the default ordering in the RayStation treatment planning system 11B (RaySearch Laboratories, Stockholm, Sweden).
\begin{algorithm}
\caption{A template for the two heuristics:  \textit{2-opt} and \textit{place after}}\label{alg:heur}
\begin{algorithmic}

\Procedure{heuristicTemplate}{spots, order, increment}
\State{\Comment{The increment parameter controls how coarse the resolution of the objective function is.}}
\State{bestFlashCoverage := calulateFlashCoverage(spots, order, increment)}
\State{anyChange := False}
\State{\Comment{The order of the outer for-loop is shuffled to make sure that the algorithm return different results when restarted.}}
\State{shuffle(spots)}
\For{s1 in spots}
\State{coveragesAndOrders := \{\}}
\State{\Comment{Note that this inner for-loop can be run in parallel.}}
\For{s2 in spots/\{s1\} }
\State{newOrder := heuristic(order, s1, s2)}
\State{flashCoverage := calulateFlashCoverage(spots, newOrder, increment)}
\State{coveragesAndOrders.add([flashCoverage, order])}
\EndFor
\State{maxCoverage, orderOfMaxCoverage := max(coveragesAndOrders)}
\If{maxCoverage $>$ bestFlashCoverage}

\State{bestFlashCoverage := flashCoverage}
\State{order := orderOfMaxCoverage}
\State{anyChange := True}

\EndIf

\EndFor
\State{\Return order, anyChange}
\EndProcedure
\end{algorithmic}
%\caption{The algorithm template for the three heuristics: 2-opt, place after and swap.}
\end{algorithm}

\subsection{Algorithms}
\subsubsection{Main algorithm}
%SAFFIRA (Spot Arrangement For FLASH In Radiotherapy Applications)
The main novelty of this article is an ad-hoc TSP heuristic. The objective function of the TSP heuristic is customized to achieve FLASH in as many voxels as possible instead of the default TSP objective of finding the tour of shortest length. Our goal was that the algorithm should be able to increase the FLASH effect of a plan in a matter of minutes or even seconds, without altering the dose distribution.

An objective function that can accomplish this is the following, which counts the number of voxels achieving FLASH:
\begin{equation}\label{eq:binary}
\sum_{v \in V} H(r_v - \hat r),
\end{equation}
where $V$ is the set indexing voxels with dose above the dose criterion level (i.e., 4 Gy in this paper) in structures where the FLASH effect is desired, $r_v$ is the dose rate in voxel $v$, $\hat r$ is the threshold dose rate, and $H$ is the Heaviside step function. In summary, given a region of interest, we strive to optimize the number of voxels inside this region that fulfill the FLASH conditions. This is equivalent to optimizing the volume of the FLASH treatment, since all voxels have equal volume.

The binary nature of objective~\eqref{eq:binary} however is not ideal for optimization; if there are no voxels fulfilling the FLASH criterion in the initial spot order, then the algorithm would have to blindly guess until the dose rate reaches the threshold. Instead, we use the objective function
\begin{equation}\label{eq-objective}
\sum_{v \in V} f(r_v)
\end{equation}
where
$$
f(r_v) :=
\begin{cases}
    0.99 \frac{e^{  r_v/\hat r - 1}}{|V|},& \text{if } r_v < \hat r\\
    1,              & \text{otherwise}
\end{cases}
$$
%\sum_{v \in V} \min\left(1,  
%\frac{e^{  r_v/\hat r - 1}}{|V|}\right),

where $|V|$ is the number of elements in $V$. This objective function gives the optimizer an incentive to increase the dose rate even if no voxel passes the threshold $S$. Moreover, a single voxel surpassing the threshold is always more important than the dose rate of all the voxels that do not surpass it, combined.  The function \textbf{calulateFlashCoverage} in Algorithms \ref{alg:heur}, \ref{alg:inc} and \ref{alg:main} used in our experiments computes~\eqref{eq-objective}.

To increase the quality of the solution, at the expense of run time, we restart the algorithm five times and return the best solution. The heuristic work on a pair of spots at the time, and the order of these pairs is shuffled to make sure that the runs do not all return the same result. The initial order of the actual spots is the same in all restarts: it is RayStation's default ordering.

To improve computational speed, the algorithm can be modified not to always consider all voxels in $V$, but instead compute an approximation of~\eqref{eq-objective}. A parameter called the $increment$ is passed to \textbf{calculateFlashCoverage}. All voxels are considered when the increment is one, every other voxel is considered when the increment is two, and so on. The final value is then multiplied with the increment value to approximate the number of voxels that fulfill the FLASH conditions. In the approximate version of the algorithm, the value of the increment is auto-tuned after each heuristic is run, by calculating the baseline objective value in which the increment is 1, and then increasing the increment until the difference between the objective value and baseline value becomes larger than 5\% of the baseline value. The increment is then set to the last integer that did not cause this difference to be larger than 5\%. See Algorithm \ref{alg:inc}. Note that the 5\% limit is an arbitrary trade-off between quality and speed, and is in no way related to the 5\% buffer that we employ in the dose rate calculation. We will refer to the algorithm that employs this voxel reduction as \textit{approximate}.

%\begin{algorithm}
%\caption{An algorithm with caption}\label{alg:obj}
%\begin{algorithmic}
%\Procedure{calculateFlashCoverage}{spots, order, increment}
%\State{startTimePerSpot := [0 for _ in range($|spots$)]}
%\State{previousIDx := order[0]}
%\State{cumsumTime := 0}
%\For{i in range($|spots|$}
%\If{i$>$0}
%\State{}
%\EndIf
%\EndFor
%\State{out := 0}
%\For{v=0; v $<$ nrOfVoxels; v += increment}
%\EndFor
%\EndProcedure
%\end{algorithmic}
%\end{algorithm}
\begin{algorithm}
\caption{An algorithm to reduce the number of voxels to consider}\label{alg:inc}
\begin{algorithmic}

\Procedure{updateIncrement}{spots, order}

\State {exactObj := calulateFlashCoverage(spots, order, increment)}

\If{exactObj $<$ 1}
\State{\Return 1}
\EndIf

\State {approximateObj := exactObj}
\State {increment = 1}
\While{abs(exactObj - approximateObj) / exactObj $\leq$ 0.05}
\State {increment++}
\State {approximateObj := calulateFlashCoverage(spots, order, increment)}
\EndWhile
\State{\Return increment - 1}
\EndProcedure

\end{algorithmic}
\end{algorithm}

The full spot ordering algorithm is shown in detail in Algorithm \ref{alg:main}.

The user might prioritize sparing some organs over others. Therefore, the set of relevant voxels $V$ can be limited to a certain region of interest (ROI). Note that only voxels with doses above the threshold are considered. In this paper we will investigate the effect of limiting the set of relevant voxels to three ROIs: the bladder, the rectum, and the patient minus the prostate.

\begin{algorithm}
\caption{The main algorithm}\label{alg:main}
\begin{algorithmic}
\Procedure{optimizePathForFLASH}{initialOrder, spots, numberOfStarts, approximate}

%\State{initialOrder := getFastPath(spots)}
\State{increment := 1}
\State{bestOrder := initialOrder}
\State{bestCoverage := calulateFlashCoverage(spots, initialOrder, increment)}
\For{i in range(numberOfStarts)}
\State{order := initialOrder}
%\For{i in range(M)}
\State{anyChange := True}
\While{anyChange}
\State{anyChange := False}
\For{heuristic in \{\textit{2-opt}, \textit{place after}\}}

\If{approximate}
\State {increment := updateIncrement(spots, order)}
\EndIf
\State{order, anyChangeFromHeuristic := heuristic(spots, order, increment)}
\State{\Comment{The structure of the heuristic is according to Algorithm~\eqref{alg:heur}.}}
\If{anyChangeFromHeuristic}
\State{anyChange := True}
\EndIf
\EndFor
%\If{not anyChange}
%\State{break}
%\EndIf
\EndWhile
\State{coverage := calulateFlashCoverage(spots, order, 1)}
\If{coverage $>$ bestCoverage}
\State{bestCoverage := coverage}
\State{bestOrder := order}
\EndIf
\EndFor
\State{\Return bestOrder}
\EndProcedure
\end{algorithmic}
\end{algorithm}

\subsubsection{Comparison Algorithms}
Two additional algorithms were created in order to determine a frame of reference for the run time and results. The first is a very fast algorithm that simply minimizes the delivery time using standard metric TSP solvers. It is based on RayStation's spot sorting functionality, modified to minimize beam travel \textit{time} instead of the usual beam travel \textit{distance}. Minimizing time is not the same as maximizing FLASH, but it is significantly faster to calculate since one doesn't have to recalculate the delivery time of the entire path when a small part of it is changed.

The second comparison algorithm is a slow global optimization the purpose of which is to give an estimation of the global FLASH coverage maximum, and as such it used the same objective function as our other algorithm. TSP is NP-hard~\cite{karp21}, so the actual maximum objective value is not expected to be found. The algorithm is a multi-restart simulated annealing optimization~\cite{sa}. It was allowed to run for 1,000,000 iterations with a reheating of the temperature every 100,000:th iteration. The temperature $t$ started at 1 and was reduced by a factor of $1-10^{-4} $ every iteration. The acceptance probability function was defined as $$ e^{\frac{f_{\textrm{new}} - f_{\textrm{old}}}{t}} $$where $f_{\textrm{new}}$ and $f_{\textrm{old}}$ is the new and old objective function value, respectively. The algorithm was run 25 times in parallel, with a random starting order each time, and the best solution found from all runs was presented. Every iteration the spot order was changed in one of three ways: either using the \textit{2-opt} heuristic, the \textit{place after} heuristic or a third heuristic in which two nodes change place. The inclusion of his third heuristic in our TSP-based algorithm was considered, but it increased the computational running time without any major improvements to the result. We decided to include it the global algorithm since it is the effect of a prioritization of the result over the run time.

We also compared the default spot ordering, were the beam scans one row at the time, from top to bottom. It switches direction every row.

In summary, we studied five algorithms: The default ordering, the minimized-time sorting, the global optimization, the novel TSP heuristic sorter, as well as its approximate version. 

\subsection{Treatment Delivery}
The beam model assumes independent beam scanning speed in the x-direction and y-direction. The Conformal FLASH delivery machine is modelled for the single highest energy of 228 MeV and the so-called hedgehog (see below), range shifter and aperture block are simulated explicitly in the Monte Carlo dose engine. The speed in the x-direction was 1553 cm/s and the speed in the y-direction was 333 cm/s. Each spot had a dead time of 1.61 ms and a minimum irradiation time of 1.25 ms. The meterset rate was 20826 MU/s, which corresponds to a machine current of 500 nA.% after taking the constant RBE factor of 1.1 into account.

A FLASH-specific research version of RayStation 11B was employed. The treatment plans were Conformal FLASH plans, as described in Nilsson~\cite{Rasmus}, with the difference that our plans were non-robust. Briefly, this means that a regular IMPT was created first. It was then converted to a FLASH plan, where each beam consists of a single energy layer. The dose was spread out in the beam direction by the use of a so-called hedgehog, which is a 3D printed, beam specific energy modulator consisting of lots of spikes (whence the name hedgehog). As the protons pass through such a spike the distribution of their energy will shift towards that of a spread out Bragg peak, thus creating a uniform dose. The hedgehog, a range shifter and a milled aperture block were used to ensure that the target would be uniformly covered with a sharp drop off, with only one energy layer.

\subsection{Patients}
The dataset consisted of 30 prostate patients, picked randomly from the Anatomical Edge Cases dataset~\cite{CLAUNCH2020e377}, but four patients were removed: patient 009 did not have a segmented rectum, the target in patient 084 could not be fully covered by both beams due to its depth, and the doses in patients 014 and 124 did not meet our goals (mentioned below). Each patient was to be treated with a hypo-fractionated plan with 8 fractions and a target dose of 40 Gy. The target was defined as the segmented prostate, plus a 5 mm margin. We removed plans where $D_{5\%} > 40.5$ Gy or $D_{95\%} < 38.25$ Gy in the target. Each plan had two opposite beams, and each fraction employed only one beam. This choice was made in order to increase the dose and dose rate for each beam in each fraction.

% Moved this to the above subsection /Viktor
%A FLASH-specific research version of RayStation 11B was employed. The plans were Conformal FLASH plans, as described in Nilsson~\cite{Rasmus}, with the difference that our plans were non-robust. Briefly, this means that a regular IMPT was created first. It was then converted to a FLASH plan, where each beam consists of a single energy layer. The dose was spread out in the beam direction by the use of a so-called hedgehog, which is a 3D printed, beam specific energy modulator consisting of lots of spikes (whence the name hedgehog). As the protons pass through such a spike the distribution of their energy will shift towards that of a spread out Bragg peak, thus creating a uniform dose. The hedgehog, a range shifter and a milled aperture block were used to ensure that the target would be uniformly covered with a sharp drop off, with only one energy layer.

The FLASH plan was spot filtered after 100 iterations, using a threshold of 26 MU/fraction in order avoid having any spots with a delivery time less than the minimum delivery time of 1.25 ms. Then the spot weight optimization was allowed to continue for 100 more iterations.

The parameters of our models were tuned towards two arbitrary patients separate from the 26 random patients that the results below are based on, to avoid over-fitting.

\subsection{Experiments}
We ran the five spot sorting algorithms for each beam, patient and ROI chosen for FLASH optimization and recorded the run time and the number of voxels that satisfied the FLASH conditions. The chosen ROI was either the rectum, the bladder or the entire patient except for the prostate. The choice of ROI indicates to the algorithm which voxels it should consider when optimizing the FLASH effect. All voxels inside the ROI, that receive $\geq 4$ Gy per fraction are considered in the algorithm. Each voxel was a cube with side 3 mm. All experiments were run on a 14 core Intel 19-10940X CPU.

\section{Results}
\begin{figure}
    \centering
    \includegraphics[scale=0.5]{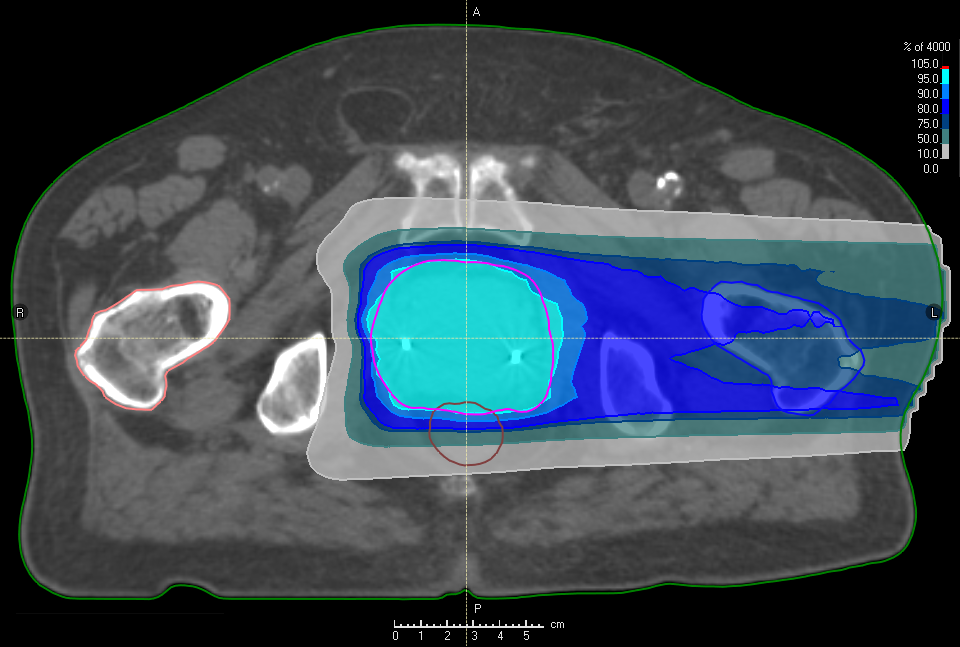}
    \caption{The dose from the left beam in a conformal FLASH plan.}
    \label{fig:82_left}
\end{figure}

\begin{figure}
    \centering
    \includegraphics[scale=0.4]{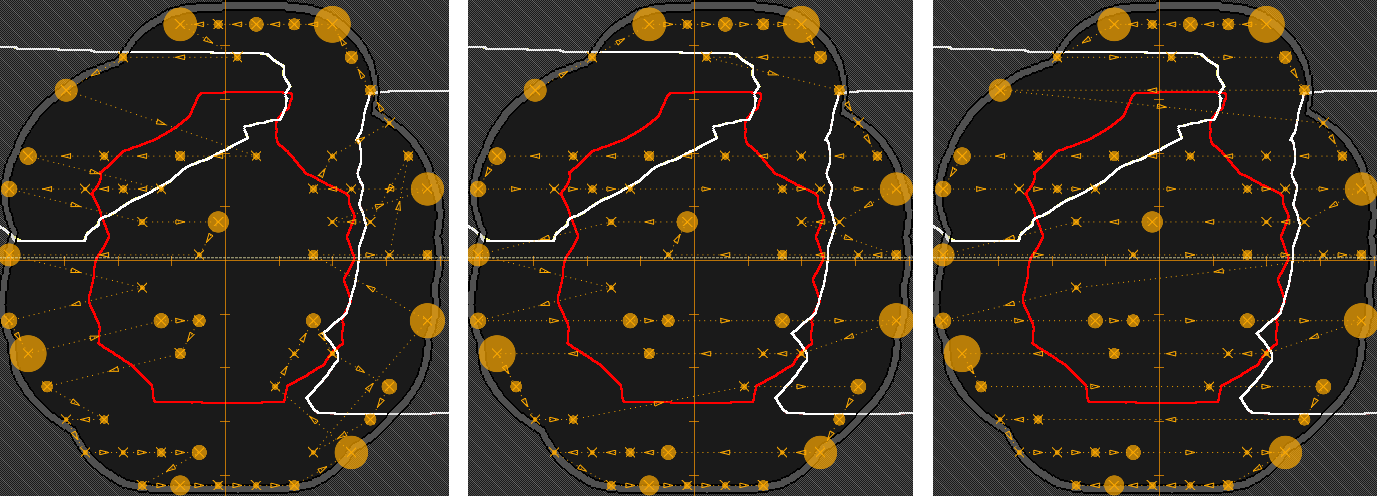}
    \caption{The spot order result for one beam for the Approximate TSP Heuristic (left), the minimized time algorithm (middle), and the default order (right). The prostate can be seen in red, with the bladder to the left and rectum to the right. The ROI selected for FLASH optimization was the entire patient minus the prostate.}
    \label{fig:patterns}
\end{figure}

\begin{figure}
    \centering
    \includegraphics[scale=0.75]{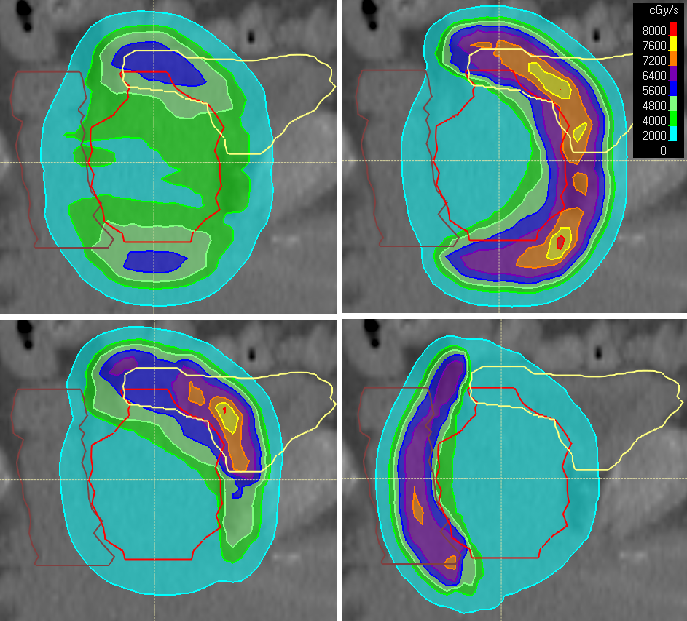}
    \caption{Sagittal view of dose rates from the left beam in a conformal FLASH plan. The top left is the default spot sorting, and the others were sorted using the approximate TSP heuristic. The top right image shows the result of an optimization of the number of FLASH fulfilling voxels in the entire patient except the prostate, the bottom left likewise but in the bladder and the bottom right in the rectum.}
    \label{fig:82_left_doserate}
\end{figure}

The dose distribution for the left beam of a representative patient can be seen in Figure~\ref{fig:82_left}, and the corresponding spot order patterns in Figure~\ref{fig:patterns}. Figure~\ref{fig:82_left_doserate} contains different dose rate distributions, showing how the high dose rate regions move between the OARs depending on where FLASH dose rates are requested from the algorithm. %, and the dose rate Volume Histogram (DRVH) can be seen in Figure~\ref{fig:drvh_82_left}. This DRVH compares the results of the default spot ordering and the ordering given by the TSP heuristic with approximate objective function. All voxels of the patient were considered in the figure, which means that the dose rate is shown even if the voxel does not reach the dose threshold condition for FLASH. Throughout this article we have assumed that the dose rate should be calculated with a 5\% buffer, but the figure includes DRVH plots for 1\% to 9\% buffers, given that the spot ordering was optimized using a 5\% buffer. This shows the robustness of the result, with respect to the size of the buffer.\\

\begin{figure}
    \centering
    \includegraphics[scale=1]{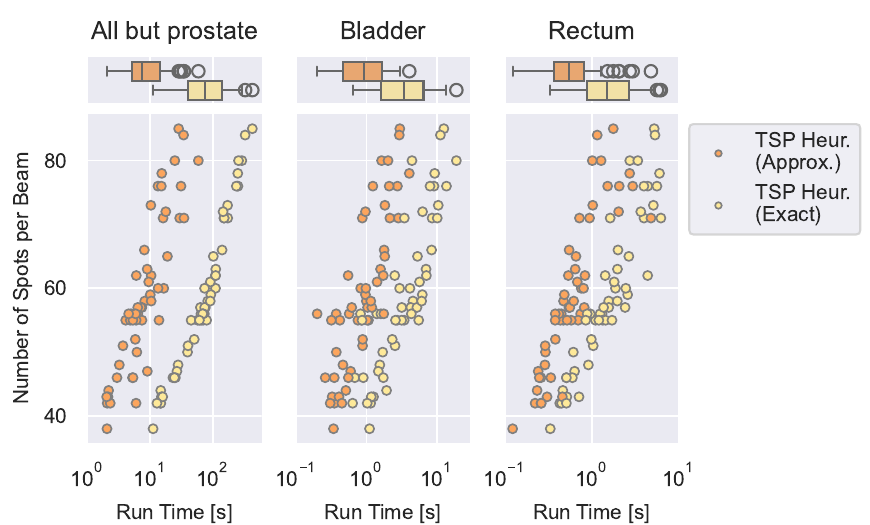}
    \caption{The computational time, per beam, for all patients and the different ROIs selected for FLASH optimization.}
    \label{fig:fast_v_slow}
\end{figure}

\begin{figure}
    \centering
    \includegraphics[scale=0.8]{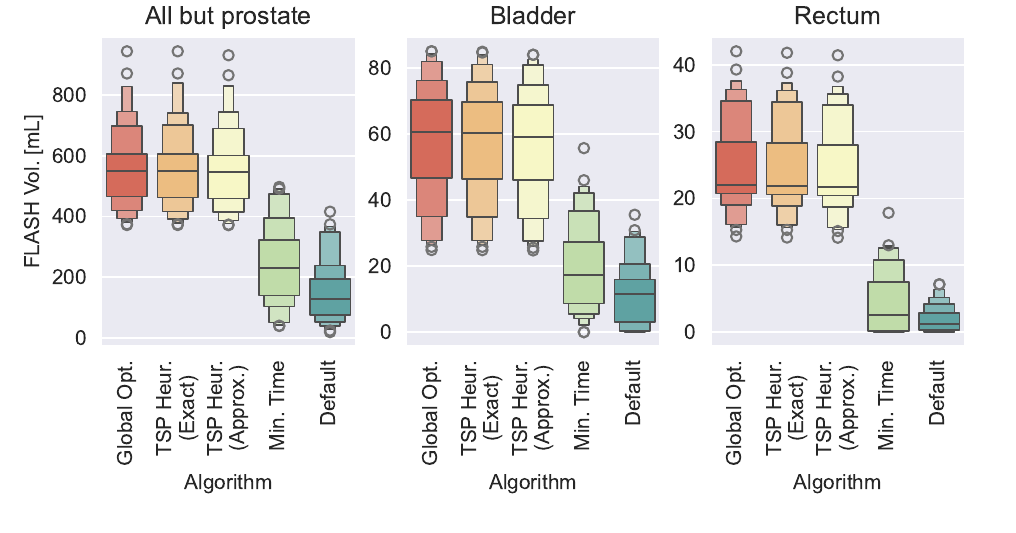}
    \caption{The distribution of volumes of the relevant ROI that satisfies the FLASH conditions, for our five tested algorithms. Note that the approximate TSP solver is almost as good as the global optimizer, even though it is more than two order of magnitudes faster.}
    \label{fig:flash_vol}
\end{figure}

%\begin{figure}
% TODO: Uppdatera med ny data
% NOTE: Jag tog bort den här plotten, för den bidrog inte med så mycket.
%    \centering
%    \includegraphics[scale=0.7]{drvh_82_left_updated_again.pdf}
%    \caption{Comparison of the distribution of dose rates for one beam of one patient, for two algorithms. The colored lines show how the distribution change as one considers different buffers in the calculation of the dose rate. The default sorting looks more robust, but novel sorting gave a larger FLASH volume at the 40 Gy/s threshold in 7 of the 9 scenarios. Note that the voxels in these graphs have not been filtered on having a dose $\geq 4$ Gy per fraction.}
%    \label{fig:drvh_82_left}
%\end{figure}
The distribution of computational times of the two TSP heuristic algorithms (exact and approximate) are shown in Figure~\ref{fig:fast_v_slow}. The approximate algorithm was substantially faster than the exact one, and the difference increased from 8 s to 18 min 15 s as the number of spots per beam increased from 39 to 115 when the chosen OAR was the full patient minus the prostate. The run time depended heavily on the ROI selected for FLASH optimization. The mean run times over the 26 patients, for each algorithm, are given in Table~\ref{tab:runtime}.\\
\begin{table}[]
    \centering
    \begin{tabular}{c|c|c|c|c|c}
            ROI& \makecell{Global\\opt.}&\makecell{TSP\\heur.\\(exact)}&\makecell{TSP\\heur.\\(approx.)}&\makecell{Minimized\\ time} & \makecell{Default}\\
           \hline
         All but prostate& 1 h 49 min 7 s & 1 min 40 s& 11.4 s & 0.7 s& 0.7 s\\
         Bladder & 5 min 45 s & 4.8 s& 1.1 s& 0.2 s& 0.2 s\\
         Rectum & 2 min 22 s & 2.1 s& 0.8 s& 0.2 s& 0.1 s
         
    \end{tabular}
    \caption{The average run times of the 26 patients for all algorithms and all ROIs selected for FLASH optimization, per beam. Note that the run time of the default sorting and minimized time sorting was dominated by the overhead time that came from collecting the relevant spotlet data.}
    \label{tab:runtime}
\end{table}
The distribution of the volumes that satisfied the FLASH criterion are shown in Figure~\ref{fig:flash_vol}, and the mean volumes are specified in Table~\ref{tab:vol}. The global optimizer yielded the largest FLASH volumes. The exact TSP heuristic resulted in $0.5\%$ smaller volumes than the global optimizer with a standard deviation of $0.9\%$. The corresponding values for the approximate TSP heuristic was $1.5\%$ with an std. of $1.0\%$. The minimized-time sorting and the default sorting resulted in substantially smaller FLASH volumes, respectively $67.5\%$ (std. $24.6\%$) and $81.3\%$ (std. $16.8\%$) lower than the global optimizer.

%The exact TSP heuristic resulted in $-2$\% to $6$\% smaller volumes than the global optimizer, and the corresponding values for the approximate TSP heuristic were $-1$\% to $7$\%. The minimized-time sorting and the default sorting resulted in substantially smaller FLASH volumes, respectively $9$\% to $100$\% and $34$\% to $100$\% lower than the global optimizer.

\begin{table}
    \centering
    \begin{tabular}{c|c|c|c|c|c|c}
            ROI& \makecell{Vol.\\$\geq 4$Gy} & \makecell{Global\\opt.}&\makecell{TSP\\heur.\\(exact)}&\makecell{TSP\\heur.\\(approx.)}&\makecell{Minimized\\ time} & \makecell{Default}\\
           \hline
         All but target & 1205 mL& 557 mL & 554 mL& 550 mL& 240 mL& 149 mL\\
         Bladder & 75 mL & 58 mL & 58 mL& 57 mL& 19 mL& 11 mL\\
         Rectum & 31 mL & 25 mL & 24 mL& 24 mL& 4 mL& 2 mL
         
    \end{tabular}
    \caption{The average volume satisfying the FLASH criteria, for all algorithms and all ROIs selected for FLASH optimization.}
    \label{tab:vol}
\end{table}

\begin{figure}
    \centering
\includegraphics[scale=0.8]{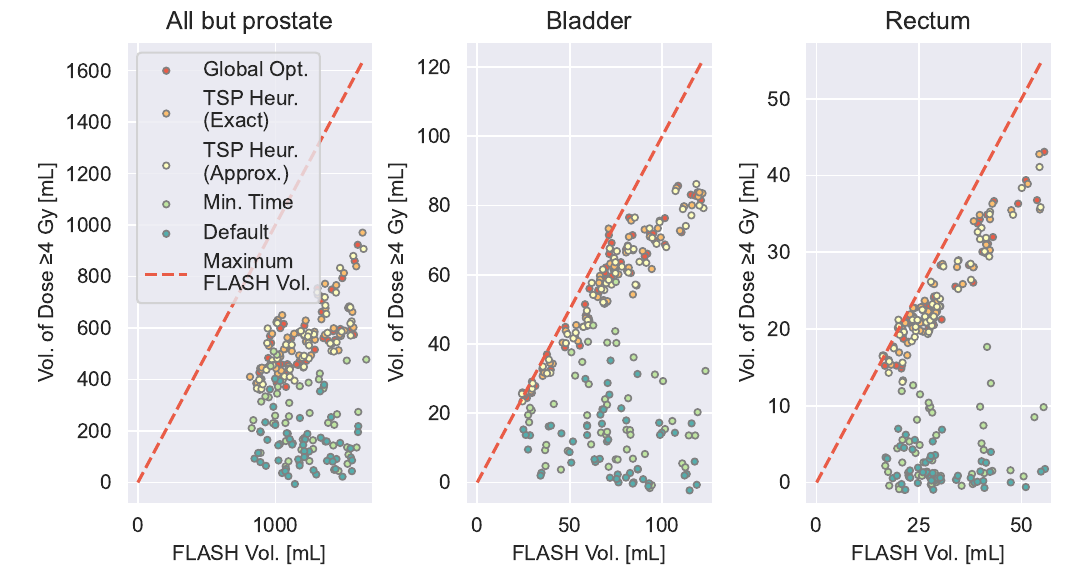}
    \caption{The volume that fulfills the FLASH criteria as a function of the volume that reaches 4 Gy. All algorithms except \textit{Min. Time} and \textit{Default} are close to the maximum reachable volume in both the Bladder cases and the Rectum cases.}
    \label{fig:scatter}
\end{figure}

\section{Discussion}
The results of our study shows that optimizing for an improved FLASH coverage can lead to substantial improvements compared to conventional spot ordering and time-based spot order optimization. More specifically, they demonstrate the potential of TSP heuristics as a tool for optimizing the FLASH effect in proton therapy. The approximate TSP heuristic sorting algorithm had a mean running time of 6 seconds per beam when the entire patient except the prostate was considered, yet its results were close to that of the global optimizer with a runtime of over an hour.

Some limitations of the present study should be noted:

The experiments were limited to employing a specific FLASH dose rate measure. Future findings might indicate that other measures correlate more strongly with the FLASH effect. We have tried to anticipate this by not relying on specifics of the dose rate measure when constructing our algorithms based on TSP heuristics. The dose rate measure can be readily replaced without changing other parts of the algorithm.

Another limitation is that the objective function employed is rather simple: maximize the number of voxels reaching FLASH. However, the algorithm can easily be generalized to contain different weights for different ROIs or voxels, or a more complicated definition of the FLASH criterion.

Furthermore, the run time grows quadratically with the number of spots, which might cause long run times for transmission (shoot-through) FLASH plans on patients with large targets.

While our method significantly reduces computational time, there is still room for further improvements and refinements. Decreasing the run time even further could enable new planning techniques, where the spot order optimization is performed simultaneously with the spot weight optimization. That way one could consider the FLASH effect while optimizing the spot weights, which might enable larger volumes receiving FLASH dose rates while maintaining high-quality dose distributions.

\section{Conclusion}
We have presented an efficient approach to optimize FLASH coverage in conformal proton therapy through the application of TSP heuristics. The results show that such heuristics can be used to achieve FLASH coverage similar to that of global optimization in a matter of seconds.

\bibliographystyle{plain}
\bibliography{main}

\begin{thebibliography}{10}

\bibitem{CLAUNCH2020e377}
C.~Claunch, A.~Kanwar, B.~Merz, S.~Rana, A.Y. Hung, and R.F. Thompson.
\newblock Anatomical edge case assessment for prostate cancer autosegmentation.
\newblock {\em International Journal of Radiation Oncology*Biology*Physics}, 108(3, Supplement):e377--e378, 2020.
\newblock Proceedings of the American Society for Radiation Oncology.

\bibitem{cancers13051012}
Shannon Cunningham, Shelby McCauley, Kanimozhi Vairamani, Joseph Speth, Swati Girdhani, Eric Abel, Ricky~A. Sharma, John~P. Perentesis, Susanne~I. Wells, Anthony Mascia, and Mathieu Sertorio.
\newblock {FLASH} proton pencil beam scanning irradiation minimizes radiation-induced leg contracture and skin toxicity in mice.
\newblock {\em Cancers}, 13(5), 2021.

\bibitem{doserateDef}
Sylvain Deffet, Valentin Hamaide, and Edmond Sterpin.
\newblock Definition of dose rate for {FLASH} pencil-beam scanning proton therapy: A comparative study.
\newblock {\em Medical Physics}, 50(9):5784--5792, 2023.

\bibitem{DIAS2015130}
Marta~F. Dias, Marco Riboldi, Joao Seco, Inês Castelhano, Andrea Pella, Alfredo Mirandola, Luís Peralta, Mario Ciocca, Roberto Orecchia, and Guido Baroni.
\newblock Scan path optimization with/without clustering for active beam delivery in charged particle therapy.
\newblock {\em Physica Medica}, 31(2):130--136, 2015.

\bibitem{NonpercentileDoseRateDef}
Michael~M. Folkerts, Eric Abel, Simon Busold, Jessica~Rika Perez, Vidhya Krishnamurthi, and C.~Clifton Ling.
\newblock A framework for defining {FLASH} dose rate for pencil beam scanning.
\newblock {\em Medical Physics}, 47(12):6396--6404, 2020.

\bibitem{10.1158/1078-0432.CCR-19-1440}
Charles Fouillade, Sandra Curras-Alonso, Lorena Giuranno, Eddy Quelennec, Sophie Heinrich, Sarah Bonnet-Boissinot, Arnaud Beddok, Sophie Leboucher, Hamza~Umut Karakurt, Mylène Bohec, Sylvain Baulande, Marc Vooijs, Pierre Verrelle, Marie Dutreix, Arturo Londoño-Vallejo, and Vincent Favaudon.
\newblock {{FLASH} Irradiation Spares Lung Progenitor Cells and Limits the Incidence of Radio-induced Senescence}.
\newblock {\em Clinical Cancer Research}, 26(6):1497--1506, 03 2020.

\bibitem{helsgaun2000effective}
Keld Helsgaun.
\newblock An effective implementation of the {L}in--{K}ernighan traveling salesman heuristic.
\newblock {\em European Journal of Operational Research}, 126(1):106--130, 2000.

\bibitem{johnson1997traveling}
David~S Johnson and Lyle~A McGeoch.
\newblock The traveling salesman problem: a case study.
\newblock {\em Local search in combinatorial optimization}, pages 215--310, 1997.

\bibitem{karp21}
Richard~M. Karp.
\newblock {\em Reducibility among Combinatorial Problems}, pages 85--103.
\newblock Springer US, Boston, MA, 1972.

\bibitem{cancers16040798}
Tyler Kaulfers, Grant Lattery, Chingyun Cheng, Xingyi Zhao, Balaji Selvaraj, Hui Wu, Arpit~M. Chhabra, Jehee~Isabelle Choi, Haibo Lin, Charles~B. Simone, Shaakir Hasan, Minglei Kang, and Jenghwa Chang.
\newblock Pencil beam scanning proton bragg peak conformal {FLASH} in prostate cancer stereotactic body radiotherapy.
\newblock {\em Cancers}, 16(4), 2024.

\bibitem{LI2015547}
Heng Li, X.~Ronald Zhu, and Xiaodong Zhang.
\newblock Reducing dose uncertainty for spot-scanning proton beam therapy of moving tumors by optimizing the spot delivery sequence.
\newblock {\em International Journal of Radiation Oncology*Biology*Physics}, 93(3):547--556, 2015.

\bibitem{liang2022investigation}
Xiaoying Liang, Chris Beltran, Chunbo Liu, Jiajian Shen, Martin Bues, and Keith~M Furutani.
\newblock Investigation of the impact of machine operating parameters on beam delivery time and its correlation with treatment plan characteristics for synchrotron-based proton pencil beam spot scanning system.
\newblock {\em Frontiers in Oncology}, 12:1036139, 2022.

\bibitem{Liu_2023}
Chunbo Liu, Chris~J Beltran, Jiajian Shen, Bo~Lu, Chunjoo Park, Sridhar Yaddanapudi, Jun Tan, Keith~M Furutani, and Xiaoying Liang.
\newblock Investigation of scan path optimization in improving proton pencil beam scanning continuous delivery.
\newblock {\em Physics in Medicine \& Biology}, 68(19):195023, sep 2023.

\bibitem{MONTAYGRUEL2017365}
Pierre Montay-Gruel, Kristoffer Petersson, Maud Jaccard, Gaël Boivin, Jean-François Germond, Benoit Petit, Raphaël Doenlen, Vincent Favaudon, François Bochud, Claude Bailat, Jean Bourhis, and Marie-Catherine Vozenin.
\newblock Irradiation in a flash: Unique sparing of memory in mice after whole brain irradiation with dose rates above 100gy/s.
\newblock {\em Radiotherapy and Oncology}, 124(3):365--369, 2017.
\newblock 15th International Wolfsberg Meeting 2017.

\bibitem{Rasmus}
Rasmus Nilsson.
\newblock A framework for creating clinically realizable robust proton conformal {FLASH} plans.
\newblock In {\em FLASH Radiotherapy and Particle Therapy Conference, Toronto}, 12 2023.

\bibitem{doserateDefPoster}
A.~Pin, D.~Tibi, L.~Hotoiu, F.~Vanderstappen, S.~Deffet, R.~Labarbe, and E.~Sterpin.
\newblock Epd012 - pencil beam proton {FLASH} therapy, field size limit with conformal {FLASH}.
\newblock {\em Physica Medica}, 94:S66--S67, 2022.
\newblock Abstracts of the FLASH Radiotherapy and Particle Therapy Conference.

\bibitem{genetic}
Rodrigo~Jos{\'e} Santo, Steven~JM Habraken, Sebastiaan Breedveld, and Mischa~S Hoogeman.
\newblock Pencil-beam delivery pattern optimization increases dose rate for stereotactic {FLASH} proton therapy.
\newblock {\em International Journal of Radiation Oncology* Biology* Physics}, 115(3):759--767, 2023.

\bibitem{dadr}
Steven van~de Water, Sairos Safai, Schippersm~Jacobus M., Damien~C. Weber, and Antony~J. Lomax.
\newblock Towards {FLASH} proton therapy: the impact of treatment planning and machine characteristics on achievable dose rates.
\newblock {\em Acta Oncologica}, 58(10):1463--1469, 2019.
\newblock PMID: 31241377.

\bibitem{vozenin2019advantage}
Marie-Catherine Vozenin, Pauline De~Fornel, Kristoffer Petersson, Vincent Favaudon, Maud Jaccard, Jean-Fran{\c{c}}ois Germond, Benoit Petit, Marco Burki, Gis{\`e}le Ferrand, David Patin, et~al.
\newblock The advantage of {FLASH} radiotherapy confirmed in mini-pig and cat-cancer patients.
\newblock {\em Clinical Cancer Research}, 25(1):35--42, 2019.

\bibitem{widenfalk2023proton}
Oscar Widenfalk.
\newblock Proton radiotherapy spot order optimization to maximize the {FLASH} effect.
\newblock Master's thesis, Uppsala University, 2023.

\bibitem{ZHAO2024}
Xingyi Zhao, Sheng Huang, Haibo Lin, J.~Isabelle Choi, Kun Zhu, Charles~B. Simone, Xueqing Yan, and Minglei Kang.
\newblock A novel dose rate optimization method to maximize ultrahigh-dose-rate coverage of critical organs at risk without compromising dosimetry metrics in proton pencil beam scanning flash radiation therapy.
\newblock {\em International Journal of Radiation Oncology*Biology*Physics}, 2024.

\bibitem{sa}
V.~Černý.
\newblock Thermodynamical approach to the traveling salesman problem: An efficient simulation algorithm.
\newblock {\em JOURNAL OF OPTIMIZATION THEORY AND APPLICATIONS}, 45(1):41--51, 1985.

\end{thebibliography}

\end{document}